\documentclass[twocolumn,twoside,slac_two]{revtex4}
\usepackage{graphicx}
\usepackage{fancyhdr}
\usepackage{hyperref}
\hypersetup{
    colorlinks=true,
    urlcolor=magenta,
}
\begin{document}

\title{The core collapse supernova rate from 24 years of data of the Large Volume Detector}

%

\author{G. Bruno, A. Molinario}
\affiliation{INFN-Laboratori Nazionali del Gran Sasso, via G. Acitelli 22, 67100 Assergi (AQ), Italy} 

\author{W. Fulgione}
\address{INFN-Laboratori Nazionali del Gran Sasso, via G. Acitelli 22, 67100 Assergi (AQ), Ital \&  Osservatorio Astrofisico di Torino, Strada Osservatorio 20, 10025, Pino Torinese (TO), Italy}
\author{C. Vigorito\footnote{e-mail:Carlo.Vigorito@to.infn.it} on behalf of the LVD Collaboration}
\affiliation{$^3$ Dipartimento di Fisica Universit\`a di Torino \&  Istituto Nazionale di Fisica Nucleare, via  P. Giuria 1, 10125 Torino (TO), Italy}

\begin{abstract}
The Large Volume Detector (LVD) at INFN Laboratori Nazionali del Gran Sasso, Italy is a 1 kt liquid scintillator neutrino observatory 
mainly  designed to study low energy neutrinos from Gravitational Stellar
Collapses (GSC) with 100\% efficiency over the entire Galaxy.
Here we summarize the results of the search for supernova neutrino bursts over the full data set 
lasting from June 1992 to May 2016 for a total live time of $8211$ days. 
In the lack of a positive observation, either in standalone mode or in coincidence with other experiments, we establish the upper limit to the rate of GSC event in the Milky Way: $0.1$ year$^{-1}$ at 90\% c.l.
\end{abstract}

\maketitle

\thispagestyle{fancy}


\section{INTRODUCTION}
Gravitational stellar collapses are astrophysical events of great interest.
Because of the complexity of the problem, the modeling of the physical processes is still in evolution, but it is in general accepted that the role of neutrinos is critical to allow the supernova to form out of a collapse \cite{Be85}. 
The confirmed detection of the neutrino signal from the SN 1987A marked the  beginning of a new era 
in neutrino astrophysics (e.g. in    
\cite{1987Hirata,1987Bionta,1987Alekseev}) and, 
in spite of some unresolved controversies \cite{lsd} opened the way for 
a new method of investigation: {\textit{The Neutrino Astronomy}}. \\
All the experiments aiming at the detection of neutrino
bursts from core collapse supernovae have to face the
extremely low frequency of this events, one every 30-50 years \cite{diehl}, 
which implies the ability to set up detectors which last in stable operation 
several years
with a very high duty cycle.

\section{THE LVD DETECTOR}
\begin{figure*}[t]
  \centering
  \includegraphics[width=\textwidth,height=80mm]{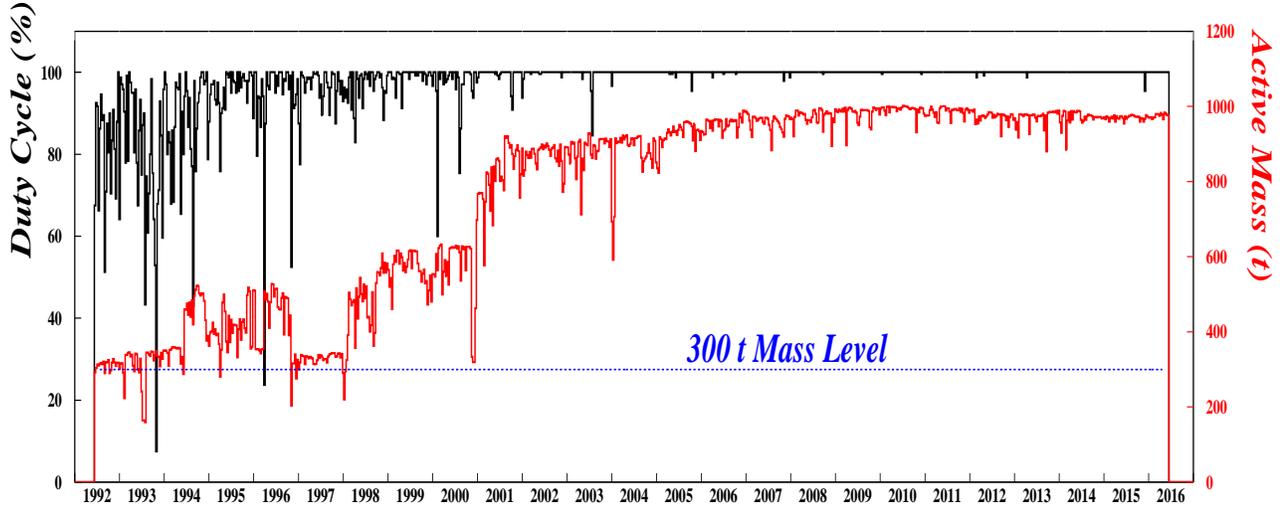}
  \caption{Duty cycle (in black) and active mass (in red) as a function of time, updated to  May, 27$^{th}$ 2016.}
  \label{mass}
\end{figure*}

LVD \cite{1992Aglietta}, located at the depth of 3600\,m w.e. in the hall A of the INFN Gran Sasso Underground Laboratory, 
consists of an array of 840 scintillator counters, 1.5 m$^{3}$ each, viewed from the top by three photomultipliers (PMTs), arranged in a modular geometry. 
The modularity of the detector results in a {\it dynamic} active mass M$_{act}$ and allows dedicated maintenance 
during running operations.
The experiment has been in operation since 1992, June 9$^{th}$ after a short commissioning phase,
its mass increasing from 300 t to its final one, 1000 t, at time of building phase completion in January 2001.
Duty cycle and active mass along the experiment life, up to 2016, May 27$^{th}$  are shown in Fig. \ref{mass}.\\
Neutrinos can be detected in LVD through charged current (CC) and neutral current (NC) interactions on proton,
carbon nuclei and electrons of the liquid scintillator (1000 t). Also interactions on iron nulcei of the
structure (850 t) may give a signal since the byproducts can reach the scintillator and be detected (see table \ref{int}). 
\begin{table}[t]
    \begin{center}
      \caption{The $\nu$ interaction channels in LVD.}
      \begin{tabular}{lccc}
        \hline
        & $\nu$ interaction channel & $\mathrm{E_{\nu}}$ threshold & $\mathrm{\%}$ \\
        \hline
        1 & $\bar \nu_\mathrm{e} + \mathrm{p \rightarrow e^{+} + n}$ & (1.8 MeV) & (88\%) \\
        2 & $\nu_\mathrm{e} + ^{12}\mathrm{C} \rightarrow ^{12}\mathrm{N} + e^{-}$ & (17.3 MeV) & (1.5\%)\\ 
        3 & $\bar \nu_\mathrm{e} + ^{12}\mathrm{C} \rightarrow ^{12}\mathrm{B} + e^{+}$ & (14.4 MeV) & (1.0\%)\\
        4 & $\nu_\mathrm{i}~ + ^{12}\mathrm{C} \rightarrow \nu_{\mathrm{i}} + ^{12}\mathrm{C}^{*}+ \gamma$ & (15.1 MeV) & (2.0\%) \\
        5 & $\nu_\mathrm{i} + e^{-} \rightarrow \nu_{\mathrm{i}} + e^{-}$ & (-) & (3.0\%)\\ 
        6 & $\nu_\mathrm{e} + ^{56}\mathrm{Fe} \rightarrow ^{56}\mathrm{Co}^{*} + e^{-}$ & (10. MeV) & (3.0\%)\\
        7 & $\bar \nu_\mathrm{e} + ^{56}\mathrm{Fe} \rightarrow ^{56}\mathrm{Mn} + e^{+}$ & (12.5 MeV) & (0.5\%) \\
        8 & $\nu_\mathrm{i}~ + ^{56}\mathrm{Fe} \rightarrow \nu_{\mathrm{i}} + ^{56}\mathrm{Fe}^{*}+ \gamma$ & (15. MeV) & (2.0\%)\\ 
      \hline  
      \end{tabular}
      \label{int}
    \end{center}
Cross sections of different interactions are obtained referring to \citep{updStrumia} for interaction 1, \citep{1988Fukugita} for interactions 2-4, \citep{1995Bahcall} for interaction 5 and \citep{2001Kolbe} and \citep{2001Toivanen} for interactions 6-8.
\vspace{-0.5cm}
\end{table}
The trigger logic is optimized for the detection of both products of the inverse beta decay (IBD) 
$\bar\nu_{e}p \rightarrow e^+ n$, the main neutrino interaction channel in LVD, 
and is based on the coincidence of the 3 PMTs of each single counter, providing a mean energy threshold of $\simeq 4$ MeV.

\section{SEARCH FOR NEUTRINO BURSTS}
\begin{figure}
  \includegraphics[width=85mm]{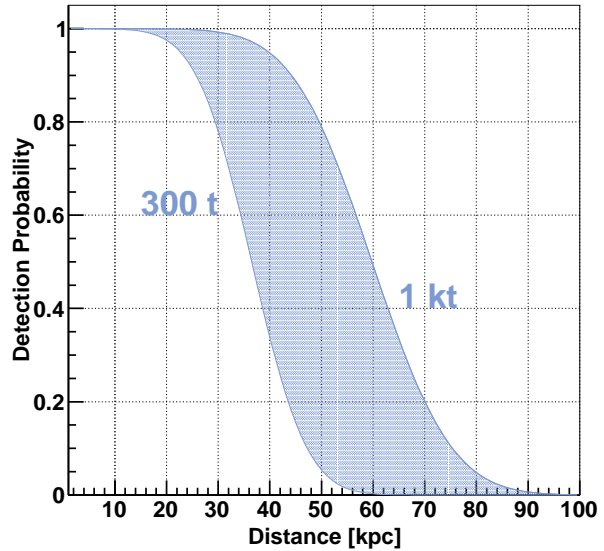}
  \caption{\label{sens} LVD detection probability versus source distance for the imitation frequency of 1/100 y$^{-1}$ (see text) for the GSC events. The light blue band corresponds to the active mass range between 300 and 1000 t.}
\end{figure}

To search for supernova neutrino bursts, we analize the time series of the selected events, i.e. triggers in the [10-100] MeV energy range ($\bar{f}_{bk}=0.03$ Hz), and look for clusters.
While to provide the SNEWS, the on-line network of running neutrino detectors \cite{snews}, with a prompt alert we use in the burst search method (on-line mode) a fixed time window ($20$ s) \cite{online}, in this analysis (off-line mode) 
we consider different burst durations up to 100 s 
as discussed in detail in \cite{offline}. 
As discussed in \cite{online,offline}, the latter method is less model dependent than the former at a cost of a
more complex procedure, which is not feasible on-line when the clusters selection
has to be quite fast.
For the off-line method the detection probability as a function of the distance of the collapse is shown in Figure \ref{sens} for the imitation frequency of 0.01 year$^{-1}$, representing the threshold for a LVD standalone alert.\\

In both cases the selection is essentially a two-step process.
In the first step, we analyze the entire time series to search for clusters of events. The rationale of the search is that
every n-th event could be the first of a possible neutrino burst. As we do not know a priori the duration of the burst, we
consider all clusters formed by the {\it n-th} event and its successive ones.
The duration of each cluster is given by the time difference $\mathrm{\Delta t}$ between the first event  and
the last one of each sequence. The advantage of the described analysis is that it is
unbiased with respect to the duration of the possible neutrino burst.
The second step of the process consists in determining if one or more among the detected clusters are neutrino bursts candidates.
To this aim, we associate to each of them (characterised by multiplicity $\mathrm{m_{i}}$ and duration $\Delta t_{i}$) a quantity that we call imitation
frequency $\mathrm{F_{im_{i}}}$. This represents the frequency with which background fluctuations can produce, by chance, clusters with multiplicity
$m \geq m_i$ and duration $\Delta t_{i}$.
 As shown in
\cite{offline}, this quantity, which depends on ($\mathrm{m_{i}, \Delta t_{i}}$), on the instantaneous background frequency, $\mathrm{f_{bk_{i}}}$
and on the maximum cluster duration chosen for the analysis, $\mathrm{\Delta t_{max}}$(100 s), can be written as:
\begin{equation}
\mathrm{F_{im_{i}} = f_{bk_{i}}^2 \Delta t_{max} \sum_{k \geq m_{i}-2} P(k, f_{bk_{i}} \Delta t_{i}) }
\end{equation}
where $\mathrm{P(k, f_\mathrm{bk_{i}} \Delta t_{\mathrm{i}})}$ is the Poisson probability to have k events in the time window $\mathrm{\Delta t_{\mathrm{i}}}$ and 
being $\mathrm{f_\mathrm{bk_{i}}}$  the background frequency.
The introduction of the imitation frequency has a double advantage. From the viewpoint of the search for neutrino bursts,
it allows us to define a priori the statistical significance of each cluster in terms of frequency. Also, it allows us
to monitor the performance of the search algorithm and the stability of the detector as a function of the imitation frequency threshold (see \cite{offline}).\\
By analyzing the time series of 15167155 events collected in 8211 days of data-taking with M$_{act}\geq 300$ t we get 31169916 clusters 
with multiplicity $m_{i}\ge2$ and $\mathrm{\Delta t_{i} \le 100}$~s.

\begin{figure}
  \includegraphics[width=85mm]{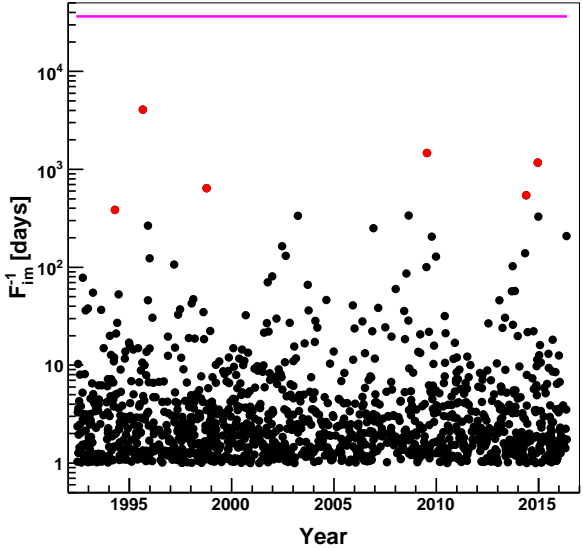}
  \caption{\label{fim} All detected clusters at $\mathrm{F_{im}}<1$ day$^{-1}$. For details see the text.}
\end{figure}

Figure \ref{fim} shows the absolute imitation frequency of all clusters
as a function of time. For convenience only clusters with $\mathrm{F_{im}}<1$ day$^{-1}$ are shown.  
It is apparent that the occurrence of clusters with different $\mathrm{F_{im}}$ over 8211 days is quite uniform.

\section{CONCLUSIONS}
None of the observed clusters passes the  0.01 year$^{-1}$ threshold (the purple line in figure \ref{fim}): 6 clusters (the red dots in the same figure) have a $F_{im}\,<\,1$ year$^{-1}$  and they have been individually checked in terms of energy spectra and low energy signals that may be the signature of the IBD  interactions. They are fully compatible with chance coincidence among background signals.\\
We conclude that no evidence of neutrino burst signal is found and taking into account the
total live-time of  LVD, 8211 days, we fix the corresponding 
upper limit to the rate of GSC out to 25 kpc of 0.1 per year at 90\% c.l.


\end{document}